\documentclass[letterpaper,twocolumn,10pt]{article}
\usepackage{algpseudocode}
\usepackage[small,compact]{titlesec}
\usepackage{comment}
\usepackage{amsmath}
\usepackage{url}
\usepackage{amssymb}
\usepackage{enumitem}
\usepackage{graphicx}
\usepackage[T1]{fontenc}
\usepackage{float}
\usepackage{amstext}
\usepackage{amssymb}
\usepackage{wrapfig}
\usepackage{times}
\usepackage{verbatim}
\usepackage{textcomp} 
\usepackage{multirow}
\usepackage{morefloats}
\usepackage{algorithmicx}

\usepackage{epsfig}
\usepackage[compact]{titlesec}
\usepackage[belowskip=0 pt,aboveskip=0 pt]{caption}

\floatstyle{ruled}
\newfloat{algorithm}{tbp}{loa}
\floatname{algorithm}{Algorithm}
\newfloat{equation}{tbp}{loa}
\floatname{equation}{Equation}
\usepackage{subfigure}

    \def\independenT#1#2{\mathrel{\setbox0\hbox{$#1#2$}%
    \copy0\kern-\wd0\mkern4mu\box0}} 
\usepackage{usenix,epsfig,endnotes}

\usepackage{usenix,epsfig,endnotes}
\begin{document}

%don't want date printed
\date{}

%make title bold and 14 pt font (Latex default is non-bold, 16 pt)
\title{\Large \bf Are Markov Models Effective for Storage Reliability Modelling?
}

%for single author (just remove % characters)
\author{
{\rm Prasenjit Karmakar}\\
Indian Institute of Science Bangalore
\and
{\rm K. Gopinath}\\
Indian Institute of Science Bangalore 
% copy the following lines to add more authors
% \and
% {\rm Name}\\
%Name Institution
} % end author

\maketitle

% Use the following at camera-ready time to suppress page numbers.
% Comment it out when you first submit the paper for review.
\thispagestyle{empty}

\subsection*{Abstract}
Continuous Time Markov Chains (CTMC) have been used extensively to
model reliability of storage systems. While the exponentially
distributed sojourn time of Markov models is widely known to be
unrealistic (and it is necessary to consider Weibull-type models for
components such as disks), recent work has also highlighted some
additional infirmities with the CTMC model, such as the ability to
handle repair times. Due to the memoryless property of these models,
any failure or repair of one component resets the ``clock'' to zero
with any partial repair or aging in some other subsystem
forgotten. It has therefore been argued that simulation is the only accurate
technique available for modelling the reliability of a storage system
with multiple components (for eg, see \cite{greenan}).

We show how both the above problematic aspects can be handled when we
consider a careful set of approximations in a detailed model of the
system. A detailed model has many states, and the transitions between
them and the current state captures the ``memory'' of the various
components. We model a non-exponential distribution using a sum of
exponential distributions, along with the use of a CTMC solver in a
probabilistic model checking tool that has support for reducing large
state spaces. Furthermore, it is possible to get results close to what
is obtained through simulation and at much lower cost.

\section{Introduction}
Traditionally, Continuous Time Markov Chains (CTMCs) have been used
%in literature 
to model RAID storage system reliability. For small systems, it is
possible to construct analytic closed-form expressions for both
transient probability of data loss as well as Mean Time To Data Loss
(MTTDL). Given some assumptions about the system, such as independent
exponential probability distributions for failure and repair, a Markov
model can be constructed, resulting often in a nice, closed-form
expression. A major problem with this model is that the reliability
calculation depends on an extremely simple view of the storage system,
especially time independence and the use of reliability models based
on exponential probability distributions. Due to the memoryless
property of these models, any failure or repair of one component
resets the ``clock'' to zero with any partial repair or aging in some
other subsystem forgotten. Hence, simulation has been argued to be
the only way to model storage reliability. While
individual simulation runs can be fast, simulation for rare
events in reliability studies requires many runs to reduce
the variance of the results (proportional to $1=\sqrt{p}$, $p$ being
the rare event probability) and techniques such as importance
sampling have to be used. However, many of its
techniques are not easy to use and are still a research topic

In this paper, we show how this problematic aspect of Markov models can be handled when we
consider a careful set of approximations in a detailed model of the
system. A detailed model has many states, and the transitions between
them and the current state captures the ``memory'' of the various
components. We show that with proper approximation of
non-exponential distributions with exponential ones, it is possible to
accurately model storage reliability using Markov models and get the same
results as simulation but much faster. We use a tool named PRISM where each module is written indep and the 
tool does the interleaving of events, so that much simpler and scalable for
programmers/designers (need to write this sentence properly).%This  We propose this method as a
%solution for systems where we can approximate non-exponential
%distributions using exponentials.
 
\section{Problem with Markov Model: Memorylessness} 
\begin{figure}[htp]
\centering
\includegraphics[height=1.0in]{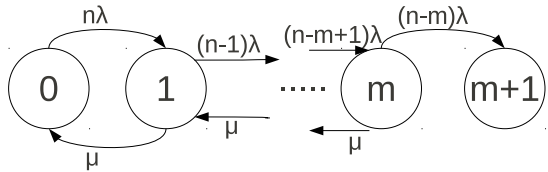}
\caption{Traditional $m$ disk fault-tolerant Markov model. Taken
  verbatim from \cite{greenan}.}
\label{multiftol}
\end{figure}
\begin{figure}[htp]
\centering
\includegraphics[height=1.0in]{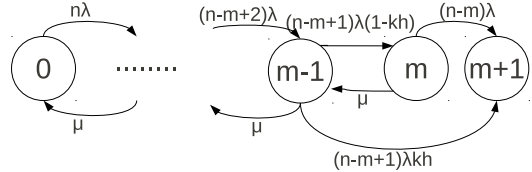}
\caption{Multi-disk fault tolerant Markov
model with latent sector errors. Verbatim from \cite{greenan}. {\it k} and
  {\it m} are the number of data disks and parity disks. {\it h} is the BER (bit-error rate) multiplied by the capacity of the device; i.e., the likelihood that a single disk exhibits a bit error if read in its entirety.}
\label{multiftolsec}
\end{figure}

%Kevin Greenan in his Phd thesis  has raised 
Several questions have been raised\cite{greenan} regarding suitability of Markov models as a tool to 
measure storage reliability. The memorylessness assumption 
made in these models may affect reliability analysis of a real system in 
case of multi-disk fault tolerant systems. To make the paper
self-contained, we consider the same Markov model
for a multi-disk fault tolerant system as in \cite{greenan} (with Figures
\ref{multiftol}, \ref{multiftolsec} taken verbatim) and summarize the insights in that
paper below.

With every failure, the system (Figure \ref{multiftol}) transitions to
a new state but where all the components in the system are reset. In
other words, the age of a still functioning available component is
reset to 0 (i.e., it becomes new), while any repair of failed
components is forgotten. Both cases are problematic. Furthermore,
consider a repair that is represented by the transition from state 1
to state 0.  Note that the repair of one disk converts all disks into their
fresh states. However, only the recently repaired component is new,
while all the others have a nonzero age.

%When the system transitions to a new state, it is as if all
%available disks are refreshed to a ``good-as-new'' state. 
%In particular, 

Next, consider the system under repair in an intermediate state $i$
with $1 < i < m$. On a failure to state $i + 1$, any previous rebuild
is lost, and only the variable $\mu$ now decides the repair
transition back to state $i$. The most recent failure therefore
determines the repair transition but it is the earliest failure, whose
rebuild is nearest to the finish, that should decide repair
transitions.

With the memorylessness assumption, therefore, each transition discards
any work completed in a previous state; hence both
component wear-out and rebuild progress are not
modelled. Such time-dependent aspects are quite difficult to
model. Furthermore, according to the analysis in \cite{greenan}, there
are differing notions of time: absolute and relative.
%The difficulty lies in the distinction between absolute time and
%relative time
Absolute time is the time since the start of the system, whereas relative times apply to individual
device lifetimes and repair clocks. Since Markov analytic models
operate in absolute time, it is not clear how to handle each individual
clock. According to Greenan, simulation is therefore the only
effective solution to this problem because simulation methods can
track relative time and thus can effectively model reliability of a
storage system with time-dependent properties.

Next consider latent sector errors. Any sector
error or bit error during rebuild in critical
mode can lead to data loss; in a $m$-disk fault tolerant system, the storage system enters
critical mode upon the $m$-th disk failure. The transition
in the Markov model in Figure \ref{multiftolsec} from the $m - 1$
to the $m + 1$ state models data loss due to
sector errors in critical mode. However, such a model
overestimates the system unreliability. A sector failure
only leads to data loss if it occurs in the portion of the
failed disk that is critically exposed. For example, in a
two-disk fault tolerant system, if the first disk to fail is
90\% rebuilt when a second disk fails, only 10\% of the
disk is critically exposed. %Figure \ref{crit} illustrates this pointin general. 
This difficulty with Markov models again follows
from the memorylessness assumption. 
%\begin{figure}[htp]
%\centering
%\includegraphics[height=1.5in]{./fig/crit.eps}
%\caption{Critical region of first failed disk susceptible to
%data loss due to latent sector errors. Verbatim from \cite{greenan}.}
%\label{crit}
%\end{figure} 

\section{Effectiveness of Markov Models}
In this paper, we argue that Markov Models are effective in spite of
the problems mentioned above; however, this requires using larger
state space models. It has been shown that it is possible to
approximate many common distributions using a sum of many exponential
distributions\cite{phasetype}; it has been computationally difficult
in the past however. Given the maturity of CTMC solvers available in
tools such as PRISM \cite{prism} and its focus on reducing the size of
state space, the difficulty is no longer an issue as we show below. To
show the effectiveness of this approach, we first show how the
reliability of RAID5 can be computed in much faster time than
simulation where disk failure is modelled by Weibull distributions. 

To handle the incorrect assumption of
time independence with respect to rebuild times, note that a 
a detailed model has
many states, and the transitions between them and the current state
captures the ``memory'' of the various components; this enables us to
avoid the time independence in large measure. We present our results
of modelling rebuild times 
in Section\ref{case2} and this agrees with simulation results
reasonably closely but at a much lower cost in terms of time and effort.

%First we present a simple example of tOur solution using Weibull approximation}
\subsection{Case Study 1: Analysis of RAID5 reliability using 3-state
  Approximation of Weibull Models}

Elerath et al. presented a sequential Monte Carlo simulation method,
using Weibull failure models, 
to calculate DDF($t$) for RAID systems where DDF($t$) is the number
of double disk failures in time $t$.
%Elerath also created a simple DDF(t) equation ~\cite{ddf_eqn} for N+1 RAID
%systems to calculate expected number of double disk failures. 
A DDF occurs when any two disks of a RAID5 group experience operational failure or one disk has a latent defect followed 
by operational failure from another disk. As PRISM does not 
support anything other than exponential distributions, we approximate Weibull distributions using phase type distributions (sum of exponentials). 
 We use the same 3 state model (burn-in, normal op, failure due to age) of \cite{infant_mortality} to
 approximate each 
of the Weibull models and find the parameters of the models $\alpha,
\sigma, \beta$ using the standard 
technique of moment matching. Here $\alpha$ is the failure rate during
burn-in, $\sigma$ the rate to working state after burn-in and $\beta$ the failure rate after burn-in. 

The pdf (probability density function) of the fail state in the 3-state
model is:  
$$\dfrac{1}{\sigma+\alpha-\beta} [\beta \sigma e^{-\beta t} + (\alpha-\beta)(\sigma+\alpha)e^{-(\sigma+\alpha)t}]$$
The first three moments of this distribution are :
\begin{align}
\mu_1&=\frac{\frac{\sigma}{\beta}+\frac{\alpha-\beta}{\alpha+\sigma}}{\alpha-\beta+\sigma}~~~~ 
\mu_2=\frac{\beta^2+\sigma(2 \alpha+\sigma)}{\beta^2 (\alpha+\sigma)^2}\\
\mu_3&=\frac{2 \left(1+\frac{-\alpha^3+\beta^3}{(\alpha+\sigma)^3}\right)}{\beta^3}.
\end{align}
Solving these three equations,  we obtain $\sigma$, $\alpha$ and $\beta$ (eqn. \ref{eqn90}).
\begin{equation}
$\sigma = \frac{4 \mu_1^3-6 \mu_1 \mu_2+\mu_3 \pm \sqrt{X}}{Y}~~ 
\alpha = -\frac{2 \left(\mu _1^3-3 \mu _1 \mu _2+\mu _3\right)}{Y} \\
\beta =  \frac{2 \mu _1^3- \mu _3 \mp \sqrt{X}}{Y} \\
X = -2\mu_1^6+6 \mu_1^4 \mu_2-18 \mu_1^2 \mu_2^2+18 \mu_2^3+8 \mu_1^3 \mu_3 -
     12 \mu_1\mu_2\mu_3+\mu_3^2 \\
Y = \mu_1^4+3 \mu_2^2-2 \mu_1 \mu_3 $
\caption{}
\label{eqn90}
\end{equation}

We use the detailed disk reliability 
model of Elerath et al. ~\cite{disk3}. %The model assumptions were as follows:
Here {\it Time to operational failure} (TTOp) (``whole disk failure'') is
modelled with a 2-parameter Weibull (shape = 1.12,  scale = 461386
hrs) whereas {\it Time to latent defect} (TTLd) is modelled as an
exponential distribution (equivalent to a Weibull with shape = 1) with scale = 9259 hours. The 
{\it Time to restore} (TTR) or rebuild time has a 3-parameter Weibull (shape  = 2,
scale = 12 hours and offset 6 hours) while
{\it Time to scrub} (TTScr) has a 3-parameter Weibull (shape  = 3,
scale = 168 hours and offset 6 hours). All of the above Weibull failure/repair models have increasing failure rates.

We equate the above moments of the 3-state model with the first three moments of Weibull for each of the three cases: TTOp,  TTScr,  TTR. For TTOp, 
the solutions turn out to be 
$\alpha = 1.72E-6$ and either 
 $\sigma = 2.49E-6$,  $\beta = 2.88E-6$ or,  equivalently
 $\sigma = 1.16E-6$,  $\beta = 4.21E-6$
%Note that due to the symmetry in the pdf, both these two solutions are
%equivalent and each can be got from the other. % by interchanging $\beta$ with $\alpha+\sigma$. % as would be expected from  because 
%they are two distinct parametrization of the same pdf.
%To check how well this pdf approximates Weibull distribution, we compare the pdf
%and hazard functions of approximate and Weibull models respectively
%(Figs.\ref{fig10:a} and \ref{fig10:b}).
\subsubsection{Comparison of Approx. Model with Weibull} %imate
\begin{figure}[htp]
  \centering
  \includegraphics[width=2.7in]{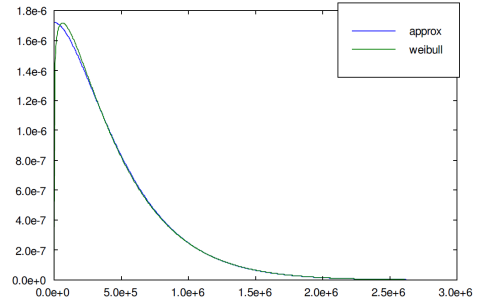}
  %\subfigure[Hazard functions]{\label{fig320 :b}\includegraphics[width=2.3in]{./fig/moment_match_hazard.eps}}
\caption{\scriptsize{Approximate vs. Weibull pdf; X axis shows time in hrs}}
\label{fig320:a}
\end{figure}

\label{sec 5.1.1}
To check how well this pdf approximates Weibull distribution, we compare the pdf functions of approximate and Weibull models 
(Figures \ref{fig320:a}). The hazard rate for the approximate model becomes constant 
after some time. This can be understood by looking into the slope of the hazard rate function for the approximate model:
$$\frac{\sigma(\beta-\alpha)e^{-(\sigma+\alpha+\beta)t}}{(\frac{\sigma}{\sigma+\alpha-\beta}e^{-\beta t}+
\frac{\alpha-\beta}{\sigma+\alpha-\beta}e^{-(\alpha+\sigma)t})^2}$$
Note that the slope function is a non-negative decreasing function for $\beta > \alpha$. Hence after some time slope becomes zero. 
%\begin{figure*}
%\includegraphics[height=3.0in]{./fig/moment_match_cdf.eps}
%\caption{CDF differences between approximate and weibull; X axis shows time in hrs}
%\label{fig33}
%\end{figure*}
 
To understand the differences better, we look at the differences between the two CDFs (Approximate minus Weibull). 
The difference is never more than +0.006 or less than -0.003.  Therefore, when using the CDFs to compute probabilities of any interval,
 the results will never be erroneous by more than 0.006 - (-0.003)  = 0.009, less than 1\%. The differences in the right tails apparently 
become zero, indicating the approximation to be very good for right tail probabilities.
%\begin{figure}[htp]
%  \centering
%  \subfigure[Pdf functions ]{\label{fig10:a}\includegraphics[width=3in]{./fig/moment_match_pdf.eps}}\hfill
%  \subfigure[Hazard functions]{\label{fig10:b}\includegraphics[width=3in]{./fig/moment_match_hazard.eps}}
%\caption{Approx. vs Weibull; X axis shows time in hrs}
%\label{fig10}
%\end{figure}
\begin{figure}[htp]
  \centering
  %\subfigure[CDF differences between approximate and weibull; X axis shows time in hrs]{\label{fig32 :a}\includegraphics[width=3.0in]{./fig/moment_match_cdf.eps}}\hfill
  \includegraphics[width=2.7 in]{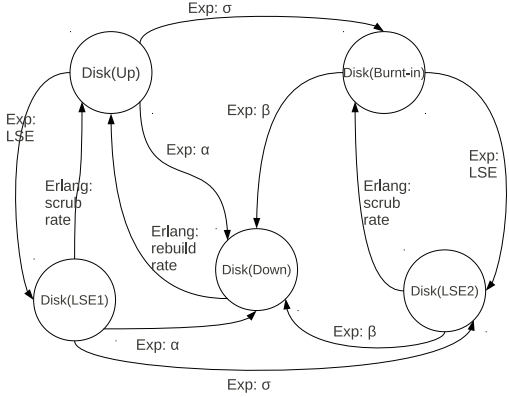}
\caption{\scriptsize{Approximate Disk model based on Gopinath et al.  ~\cite{disk2}: one
  difference is that we consider here a more accurate model that has a transition from Disk(LSE1) state to the 
Disk(LSE2) state with rate $\sigma$ rather than a transition from Disk(LSE1) to Disk(Burnt-in) state.}}
\label{fig32 :b}
\end{figure}
%\begin{figure}[htp]
%\centering
%\includegraphics[height=2in]{./fig/disk_approx_model.eps}
%\caption{Approx. Disk model based on Gopinath et al.  ~\cite{disk2}: one
%  difference is that 
%LSE: latent defect rate; 
%we consider here a more accurate model that has a transition from Disk(LSE1) state to the 
%Disk(LSE2) state with rate $\sigma$ rather than a transition from Disk(LSE1) to Disk(Burnt-in) state.
%The dotted 
%line shows the change in the model of Gopi et al ~\cite{disk2}. The new modified transition is from Disk(LSE1) state to Disk(LSE2) state. 
%}
%\label{fig11}
%\end{figure}
 
For TTR and TTScr, with the same approach, we get a complex number for $\sigma$ and 
$\beta$ and negative value for $\alpha$ for each of the 
two solutions respectively. Hence,  we use other phase type distributions 
such as Erlang distributions  ~\cite{disk2}. We use a 3-stage Erlang model. For TTScr
$\lambda$ = 0.019228232 and for TTR $\lambda$ = 0.180345653. 
Using these models for each type of failure/repair we build a detailed disk model (Fig.\ref{fig32 :b}). 

%We compare the reliability of a 6 disk RAID5 subsystem using the detailed disk model with the simulation results to
%check the accuracy of this approximation. Table \ref{table9} shows that model results are close  
%to the simulation results. When we change the number of disks in a
%RAID group and the type of RAID group (such as RAID6) and then compare the model results 
%and simulation results, they too are close; due to lack of space
%we do not present them here.
\paragraph{Comparison of PRISM, Monte Carlo Simulation Results:} % and DDF(t) equation Results :}
We compare the reliability of RAID subsystems using PRISM model and Monte Carlo Simulation %and DDF(t) equation 
(Table \ref{table18} and Table \ref{table20}). % and Table \ref{table20}). 
%To calculate 
%DDF(t)  per $10^{n}$ RAID groups in PRISM we calculate probability of dataloss for a RAID group using symmetry reduction ~\cite{symmetry}
%and multiply by $10^{n}$. 
%properly
We try to keep the variance of both PRISM and Monte Carlo Simulation 
results same so that we can make a fair comparison. Hence, we 
set the termination epsilon parameter in case of PRISM and the number of 
experiments parameter in case of Monte Carlo simulation  accordingly.  
Results from Table \ref{table18} % and \ref{table20} 
(under the column with 3-state disk failure model) show that 
DDF(t) values calculated from PRISM model are similar with those of the
Monte Carlo simulation. %and DDF(t)  equation. 
Due to the 
front-overloading of our approximate pdf (compared to the actual Weibull pdf), the difference between DDF(t) values calculated 
using PRISM and Monte Carlo simulation
%the other two methods (MC-Sim and DDF(t) equation) 
is much higher in the beginning. 
%The reason of front-overloading 
%is that the approximate hazard rate function becomes flat after some time $t$ but the Weibull hazard rate increases as $(\frac{t}{scale})^{(shape-1)}$ ($scale$ and $
%shape$ are parameters of Weibull distribution) and we match the 
%moment of both the distributions. Hence to balance both the distributions the hazard function (which is equal to probability density 
%function in the initial time period) of the approximate function is overloaded in the beginning. 
\begin{table*}[ht]
\begin{center}
\begin{tabular}{|c|c|c|c|c|c|} %{|c|c|c|c|c|c|c|c|c|}
\hline
 \textbf{Time(yr)} & \textbf{pDDF$_{3}$(t)} & \textbf{pDDF$_{4}$(t)}
 & \textbf{sDDF(t)} %& \textbf{eqDDF(t)} 
& \textbf{sDev$_{3}$(\%)} & \textbf{sDev$_{4}$(\%)} 
%& \textbf{eDev$_{3}$(\%)}   & \textbf{eDev$_{4}$(\%)
\\ \hline
 1                                   &	7.12    & 5.59   &  5.63   
%&   5.64 
& 26.5 & -0.72 %& 26.24 & -0.9
\\ \hline
 2                                   & 14.37    & 12.2   & 12.23   
%&   12.26 
& 17.5& -0.21  %& 17.21 & -0.46
\\ \hline
 3                                   & 21.67    & 19.26   & 19.21 
%&    19.31 
& 12.8& 0.28  %& 12.22 & -0.24
\\ \hline
 4                                   & 28.99    & 26.59   & 26.43 
%&    26.64 
&9.7  & 0.59  %&  8.82 & -0.20  
\\ \hline
 5                                   & 36.35    & 34.06   & 33.8 
%&     34.21 
&7.5  & 0.75 %&   6.26 & -0.45 
\\ \hline
 6                                   & 43.73    & 41.6   & 41.27 
%&     41.96 
& 6   & 0.8   %&   4.22 & -0.86 
\\ \hline
 7                                   & 51.13    & 49.17   & 48.79  
%&   49.87 
& 4.8 & 0.77 %&  2.53 & -1.41
\\ \hline
 8                                   & 58.54    & 56.73  &  56.36 
%&     57.91 
& 3.9 & 0.66 %&  1.09 & -2.09 
\\ \hline
 9                                   & 65.96    & 64.27  &  63.93 
%&     66.08 
&3.2 & 0.57 %&   -0.18 & -2.73
\\ \hline 
10                                   & 73.39    & 71.78  &  71.50 
%&    74.35 
& 2.7 & 0.38 %&  -1.29 & -3.46 
\\ \hline
\end{tabular}
\end{center}
\caption{\scriptsize{DDF(t)  per 1000 RAID groups for 6 disk RAID5: 
PRISM Model (pDDF$_{i}$(t))  vs. Simulation (sDDF(t)).  %vs. DDF(t)  equation (eqDDF(t)) result;
pDDF$_{i}$(t)= DDF calculated in PRISM using $i$-state disk failure model.
sDev = Deviation of PRISM results from Simulation results;
%eDev = Deviation of PRISM results from DDF(t)  equation results;
Time taken for Model Checking = \textbf{37 sec} (using 3-state model) and \textbf{4.3 min} (using 4-state model) 
 while time for Simulation = \textbf{8 min}; both PRISM and simulation
 error are 1\%. Number of states using symmetry reduction are 8280 and 33985 with 3-state and 4-state model respectively.
}}
\label{table18}
\end{table*}

\begin{table}[ht]
\begin{center}

\begin{tabular}{|c|c|c|c|}
\hline
 \textbf{Time(yr) } & \textbf{PRISM DDF(t) } & \textbf{sDDF(t) } & \textbf{sDev(\%) }\\
\hline
 1                                   &	2.26       &  1.92  &  17.7\\ \hline
 2                                   &  4.62       &  3.84   & 20.3  \\ \hline
 3                                   &  7.03       &  6.46 &  8.8 \\ \hline
 4                                   &  9.51       &  9.32 &  2    \\ \hline
 5                                   & 12.04       & 12.16 & -1\\ \hline
 6                                   & 14.63       & 14.87 & -1.6 \\ \hline
 7                                   & 17.27       & 18.24  & -5\\ \hline
 8                                   & 19.96      &  21.52 & -7.3 \\ \hline
 9                                   & 22.71      &  24.56 & -7.5 \\ \hline 
10                                   & 25.50      &  28.16 & -9.4 \\ \hline

\end{tabular}
\end{center}
\caption{\scriptsize{DDF(t)  per 1000000 RAID groups for 8 disk RAID6  : 
%PRISM Model (PRISM DDF(t))  vs. Simulation (sDDF(t))  results;
%sDev = Deviation of PRISM results from Simulation results; 
Time taken for Model Checking = \textbf{12.6 min} while time for Simulation = \textbf{26 hr}; 
PRISM error is 1\% and Simulation Error is 4\%}}
\label{table20}
\end{table}

It can be noted that 
the higher deviation between the results of PRISM and simulation due to front overloading of the 
approximate pdf can be reduced by adding more states in the Markov model. We consider a 4-state model 
to check how well it approximates Weibull. 
Note that a 4-state Markov model has 5 model parameters. To estimate them using 
moment matching is hard; we estimate the parameters by %trial and error method.
% i.e. we try to 
matching the hazard rate curve of approximate distribution and Weibull
distribution for some time period of interest (0 to 10 yr). Note that
the 4-state model does not have an obvious interpretation as the
3-state does. (we need to reword it. We can say we tried free tools avlbl 
but they were not upto it. Instead of developing another tool, we found it easier to try it by hand).

%The hazard rate curve and pdf. of the fail state using 4-state model, 
%Weibull model and 3-state model are shown in Figure \ref{fig4state}. Note that using a 4-state model and estimating parameters 
%using trial and error method we are able to reduce the front-overloading of approximate model. 
%Next, we check how this 4-state model performs when modelling disk subsystems. 
Table \ref{table18} (under the column 4-state disk model) shows the DDF(t) values 
computed using the 4-state model and how they agree with simulation. 
%and DDF(t) equation results. 
Note that in the time period 
of $t$ = 0 to 10 yr, the deviations are now much less (especially in
the initial period). %, but the hazard rate function starts to flatten 
 \begin{figure}[htp]
  \centering
 % \subfigure[Simulation results \cite{Greenanth}]{\label{greenan :a}\includegraphics[width=3.0 in]{./fig/greenan.eps}}\hfill
  %\subfigure[PRISM results]{\label{greenan :b}\includegraphics[width=3.0 in]{./fig/prism_greenan.eps}}
\includegraphics[width=3.0 in]{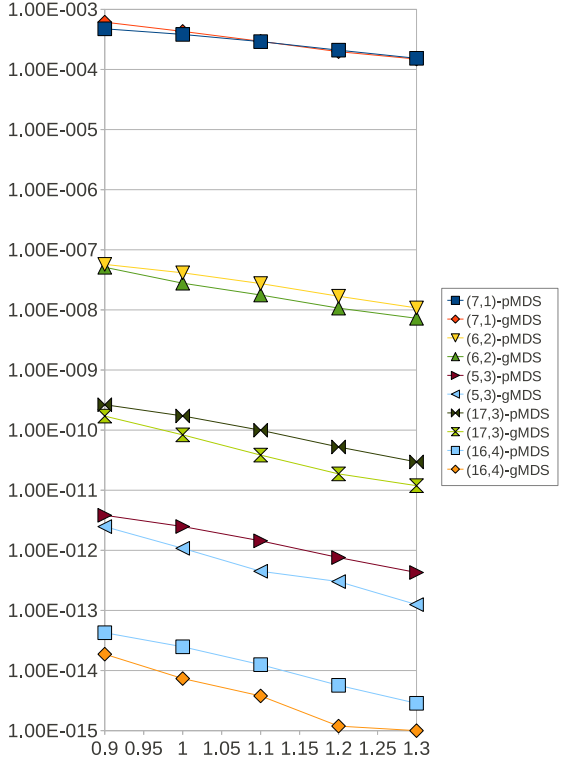}
\caption{Comparison of PRISM results with Greenan's simulation
  results. $X$ and $Y$ axis corresponds to failure shape and the probability of dataloss respectively. $(n,k)$-pMDS and $(n,k)$-gMDS corresponds to 
the result using PRISM and Greenan's simulation result respectively. The number of states is less than 1000 for PRISM (we make use of
  its symmetry reduction capability).}
\label{greenan}
\end{figure}
\subsection{Case Study 2: Comparison with Greenan's simulation results
for rebuild}
\label{case2}
For a single disk fault tolerant system, the difficulties of modelling
rebuild with Markov models does not arise. In case of multi-disk fault tolerant system (for example RAID6), we 
compose detailed disk models (Fig.\ref{fig32 :b}) to build a disk subsystem model. Hence we consider failure and repair modes of each 
disk separately rather than considering a system level Markov model like Figures \ref{multiftol} and \ref{multiftolsec}. Moreover, 
when we approximate Weibull repair and Weibull failure by summation of exponentials (i.e. by adding multiple states and transitions 
corresponding to a single failure/repair transition) then these states keep information regarding repair progress and age of a component 
respectively. Hence, our disk subsystem models using detailed disk models reduce the chance of 
loss of information due to memorylessness property significantly.

To show that Markov models are effective, we use PRISM to model some disk subsystem
configurations that use MDS (maximum distance separable) codes from Greenan's thesis \cite{Greenanth}, and 
compare PRISM results with the Greenan's simulation results from a
``high-fidelity simulator'' developed only for this purpose. 
%Note that it is a difficult task to do because all the simulation results 
% have been plotted by Greenan rather than providing the values, still
% we try to compare both the results from the plot. 
Here, different %We model the 
MDS configurations %of Section 6.4 of Greenan's thesis where he 
are analyzed to compute the sensitivity of probability 
of data loss in 10 years to failure shape parameters.
 We approximate Weibull failure by a 3-state Markov model and Weibull repair by a 8-stage Erlang model. In both cases we estimate 
the model parameters by moment matching. For some cases where the model parameters ($\sigma$ and $\beta$) of 3-state Markov model 
 result in a complex number, we estimate the model parameters based on
 %trial and error method based on 
the solution found by moment matching.  
Our results (Figure \ref{greenan}) show that PRISM results are similar in ``order'' 
 compared to the simulation results with the advantage that it is very
 fast (time taken for calculating each data point in 
Figure \ref{greenan} is less than 1 sec in PRISM). For some cases with multi-disk fault tolerant systems PRISM results are higher than 
the simulation results. The possible reasons are 
\begin{itemize}
 \item The front overloading of the approximate pdf (in a 3-state model) w.r.t. Weibull pdf.
 \item Approximating Weibull distribution using Erlang distribution in case of repair distribution is not good because Weibull repair 
has a high shape parameter (shape=2). 
\end{itemize}
The success of our technique depends on how well we approximate Weibull distribution using exponentials. For Weibull distribution 
with high shape parameter the approximation becomes poor as the hazard function for approximate becomes flat for large $t$ whereas 
for Weibull it is an increasing function (for example, with Weibull shape=2, hazard rate increases linearly with time). 
%Intuitively, 
%the hazard rate of the Weibull approximate  becomes flat because once the probability of reaching the fail state of the Weibull approximate
% is close to 1, the probability of failure after that time becomes negligible.        
%\begin{figure}
%\centering
%\includegraphics[height=3.0in]{./fig/flat.eps}
%\caption{Hazard rate curve in a long time range}
%\label{flat}
%\end{figure}
%\begin{figure}
%\centering
%\includegraphics[height=3.0in]{./fig/cdf.eps}
%\caption{CDF difference between approximate and Weibull}
%\label{cdf}
%\end{figure}         
\section{Conclusion}
In this paper, we have shown that many difficulties due to the
memorylessness of Markov models can be handled if more detailed models
are considered. A detailed model has many states, and the transitions
between them and the current state captures the ``memory'' of the various
components. Hence, we can get good agreement with similar detailed simulated
models but at lower cost in time (for example, for rare event failure
case such as RAID6, PRISM model is almost 150 times faster than
simulation at the same accuracy). We need mature tools such as PRISM to
make such detailed Markov models feasible.
Simulation may still be the best general method but we also need to
consider that validation of
the results in a rare event simulation is non-trivial. We believe that
the automation that is possible in CTMC solvers as in PRISM (for eg, of interleaving all the failure
cases) makes it much simpler to
consider detailed models.

\end{document}